\documentclass[a4paper]{jpconf}

\pdfoutput=1

\usepackage{ifthen}
\newboolean{eng}
\newboolean{rus}
\newlength{\lengrus}

\setboolean{eng}{true}

\setboolean{rus}{true}
\setboolean{rus}{false}     

\usepackage{engrus}

\mifeng{}{\mifrus{}{\setboolean{eng}{true}}}

\makeatletter
\renewcommand{\@oddhead }{\vbox{\hbox to\textwidth{%
\mifeng{A.~A.~Chernitskii}{}\mifengrus{\hspace{13.1pc}}{}%
\mifrus{А.~А.~Черницкий}{}\hfill\thepage\strut}\hrule}}
\makeatother

\usepackage{ifpdf}

\usepackage{graphicx}

\usepackage{latexsym}
\usepackage{amsfonts}
\usepackage{amssymb}
\usepackage{amsbsy}
\usepackage{amsmath}   %

\usepackage[usenames,dvipsnames]{pstricks}
\ifpdf
 \else
  \usepackage{epsfig}
  \usepackage{pst-solides3d}
\fi

\definecolor{mcolora}{rgb}{1,0,0}
\definecolor{mcolorr}{rgb}{0,0,1}

\def\textadda#1{\textcolor{mcolora}{#1}}
\def\textrema#1{\textcolor{mcolorr}{\sout{#1}}}
\def\textadda#1{#1}
\def\textrema#1{}

\def\p{\partial}

\def\stTD#1#2{\hbox to 0em{\mathsurround=0em $\stackrel{#1}{\makebox[0pt]{} #2}$\hss} \phantom{#2}}\def\stscript#1#2{\hbox to 0em{\mathsurround=0em ${\scriptstyle\stackrel{#1}{\makebox[0pt]{} #2}}$\hss} \phantom{#2}}\def\stscriptscript#1#2{\hbox to 0em{\mathsurround=0em ${\scriptscriptstyle\stackrel{#1}{\makebox[0pt]{} #2}}$\hss} \phantom{#2}}\def\st#1#2{\mathchoice{\stTD{#1}{#2}}{\stTD{#1}{#2}}{\stscript{#1}{#2}}{\stscriptscript{#1}{#2}}}
\def\comb#1#2#3{{\mathsurround 0pt\hbox to 0pt {\hspace*{#3}\raisebox{#2}{${#1}$}\hss}}}
\def\combs#1#2#3{{\mathsurround 0pt\hbox to 0pt {\hspace*{#3}\raisebox{#2}{${\scriptstyle #1}$}\hss}}}
\def\combss#1#2#3{{\mathsurround 0pt\hbox to 0pt {\hspace*{#3}\raisebox{#2}{${\scriptscriptstyle #1}$}\hss}}}

\def\EMT{\mathchoice{\combs{\to}{0.3ex}{-0.2ex}{T}}{\combs{\to}{0.3ex}{-0.2ex}{T}}{\combss{\to}{0.2ex}{-0.2ex}{T}}{\combss{\to}{0.2ex}{-0.2ex}{T}}}

\def\q{\mathchoice{\combs{-}{0.15ex}{-0.2ex}q}{\combs{-}{0.05ex}{-0.2ex}q}{\combss{-}{0.05ex}{-0.25ex}q}{}{}}

\def\ffun{\Phi}

\def\dffun{\ffun}

\def\ffind{\Upsilon}
\def\df{\mathrm{d}}

\def\metr{\mathfrak{m}}

\def\metrp{\mathchoice{\comb{-}{-0.9ex}{0ex}\mathfrak{m}}{\comb{-}{-0.9ex}{0ex}\mathfrak{m}}{\combs{-}{-0.75ex}{-0.1ex}\mathfrak{m}}{}{}}

\def\bje{{\mathsurround 0pt\lower.0ex\hbox{${\scriptscriptstyle \mathbf{e}}$}\mspace{-3.4mu}\mathbf{j}}}
\def\je{{\mathsurround 0pt\lower.0ex\hbox{${\scriptscriptstyle e}$}\mspace{-4.5mu}j}}
\def\bjm{{\mathsurround 0pt\lower.0ex\hbox{${\scriptscriptstyle \mathbf{m}}$}\mspace{-5.6mu}\mathbf{j}}}
\def\jm{{\mathsurround 0pt\lower.0ex\hbox{${\scriptscriptstyle m}$}\mspace{-7.0mu}j}}

\def\p{\partial}

\def\Lor{L}

\def\eqdef{\doteqdot}

\def\Sur{\Sigma}

\def\dSur{{\rm d}\hspace{-0.3ex}\Sigma}

\def\Surc{\sigma}

\def\dSurc{{\rm d}\hspace{-0.3ex}\sigma}

\def\Fo{\mathbb{F}}

\def\Vols{\mathchoice{\combs{\triangle}{0.45ex}{0.1ex} {\rm V}}{\combs{\triangle}{0.45ex}{0.1ex} {\rm V}}{\combss{\triangle}{0.2ex}{0ex} {\rm V}}{}{}}
\def\dVols{\mathrm{d}\mathchoice{\combs{\triangle}{0.45ex}{0.1ex} {\rm V}}{\combs{\triangle}{0.45ex}{0.1ex} {\rm V}}{\combss{\triangle}{0.2ex}{0ex} {\rm V}}{}{}}

\def\xp#1{\comb{\cdot}{-1ex}{0.3ex}{#1}}

\def\xc{\mathchoice{\comb{\boldsymbol{\cdot}}{-0.1ex}{-0.05ex}x}{\comb{\boldsymbol{\cdot}}{-0.1ex}{-0.05ex}x}{\combs{\boldsymbol{\cdot}}{-0.05ex}{-0.05ex}x}{}{}}
\def\bxv{\mathchoice{\comb{\cdot}{0.2ex}{0.5ex}\mathbf{v}}{\comb{\cdot}{0.2ex}{0.5ex}\mathbf{v}}{\combs{\cdot}{0.15ex}{0.35ex}\mathbf{v}}{}{}}
\def\xv{\mathchoice{\comb{\boldsymbol{\cdot}}{-0.1ex}{0.3ex}v}{\comb{\boldsymbol{\cdot}}{-0.1ex}{0.3ex}v}{\combs{\boldsymbol{\cdot}}{-0.05ex}{0.2ex}v}{}{}}

\def\pu{\mathchoice{\comb{\boldsymbol{\cdot}}{-0.1ex}{0.3ex}u}{\comb{\boldsymbol{\cdot}}{-0.1ex}{0.3ex}u}{\combs{\boldsymbol{\cdot}}{-0.05ex}{0.2ex}u}{}{}}
\def\mass{{\sf m}}

\def\delf{{\mathsurround=0ex {\displaystyle\comb{\cdot}{-0.1ex}{0.07em}\delta}}}

\def\Cron{\mathchoice{\combs{\backslash}{0.1ex}{0.1ex}\delta}{\combs{\backslash}{0.1ex}{0.1ex}\delta}{\combss{\backslash}{0ex}{0ex}\delta}{}{}}

\def\Energy{\mathbb{E}}

\def\EMV{\mathbb{P}}

\def\Vol{\mathchoice{\combs{\square}{0.15ex}{0.2ex} {\rm V}}{\combs{\square}{0.15ex}{0.2ex} {\rm V}}{\combss{\square}{0.12ex}{0.095ex} {\rm V}}{}{}}

\def\xxx{\chi}

\def\metrEff{\mathchoice{\combs{\sim}{1ex}{0.2ex}\mathfrak{m}}{\combs{\sim}{1ex}{0.2ex}\mathfrak{m}}{\combss{\sim}{0.66ex}{0.05ex}\mathfrak{m}}{}{}}
\def\metrEffm1{\check{\metrEff}}

\def\HypGeomF#1#2#3{\mathchoice{\comb{\diamond}{0.25ex}{0.1em}F_{#1; #2; #3}}{\comb{\diamond}{0.25ex}{0.1em}F_{#1; #2; #3}}{\combs{\diamond}{0.2ex}{0.05em}F_{#1; #2; #3}}{}{}}

\def\br{\bar{r}}

\def\ps{\mathfrak{s}}

\def\LF{\mathchoice{\combs{-}{0.3ex}{-0.1ex}\mathcal{L}}{\combs{-}{0.3ex}{-0.1ex}\mathcal{L}}{\combss{-}{0.25ex}{-0.12ex}\mathcal{L}}{}{}}

\begin{document}

\phantom{.}
\vspace{-3.2pc}
\vbox{\hbox to\textwidth{%
Journal of Physics: Conf. Series {\bf 938} (2017) 012029\hfill doi:10.1088/1742-6596/938/1/012029\strut}\hrule
\vspace{0.6ex}
\noindent
XVII Workshop on High Energy Spin Physics "DSPIN-2017"}

\mifeng{%
\title[About long-range interaction of spheroidal solitons]{About long-range interaction of spheroidal solitons\\ in scalar field nonlinear model}
\author{Alexander A. Chernitskii}
\address{$^1$ Department  of Mathematics\\ St. Petersburg State Chemical Pharmaceutical Academy\\Prof. Popov str. 14, St. Petersburg, 197022, Russia}
\address{$^2$ A. Friedmann Laboratory for Theoretical Physics\\St. Petersburg, Russia}
\ead{AAChernitskii@mail.ru}
\begin{abstract}
The nonlinear scalar field model of space-time film (Born -- Infeld type nonlinear scalar field model) is considered.
Its spherically symmetrical solution is obtained. This solution gives the class of moving solitary solutions or solitons with the Lorentz transformation.
We consider the distant interaction between such spheroidal solitons or spherons. This interaction is caused by the nonlinearity of the model.
Starting from the static configuration with two spherons we show that the interaction under investigation is similar to electromagnetic one.
\end{abstract}
}{}

\mifengrus{\vspace{-10ex}}{}


\mifrus{%
\selectlanguage{russian}
\title[О дальнем взаимодействии сфероидальных солитонов]{О дальнем взаимодействии сфероидальных солитонов в нелинейной модели скалярного поля}
\author{Александр А. Черницкий}
\address{$^1$ Кафедра математики\\ Санкт-Петербургская Химико-Фармацевтическая Академия\\ул. Проф. Попова 14, Санкт-Петербург, 197022, Россия}
\address{$^2$ Фридмановская Лаборатория Теоретической Физики\\Санкт-Петербург, Россия}
\ead{AAChernitskii@mail.ru}
\begin{abstract}
Рассматривается нелинейная скалярная полевая модель пространственно-временной плёнки (нелинейная скалярная полевая модель типа Борна -- Инфельда).
Получено её сферически симметричное решение. Это решение порождает класс движущихся уединённых решений или солитонов при помощи преобразований Лоренца.
Мы рассматриваем дальнее взаимодействие между такими сфероидальными солитонами или сферонами.
Это взаимодействие обусловлено нелинейностью модели.
Взяв за основу статическую конфигурацию с двумя сферонами, мы показываем, что исследуемое взаимодействие подобно электромагнитному.
\end{abstract}
}{}

\mifengrus{\newpage}{}
\engrus{1ex}{1ex}{%
\section{Introduction}
\label{introd}
}{%
\mifeng{\addtocounter{section}{-1}}{}
\section{Введение}
\mifeng{}{\label{introd}}
}

\engrus{0.5ex}{0.5ex}{
We consider the nonlinear scalar field model that can be called the extremal space-time film one \cite{Chernitskii2016a,Chernitskii2015a}.
The investigation of this field model is interesting from the various point\textadda{s} of view.
}{
Мы рассматриваем нелинейную скалярную полевую модель, которую можно назвать моделью экстремальной пространственно-временной плёнки \cite{Chernitskii2016a,Chernitskii2015a}.
Исследование этой полевой модели интересно с различных точек зрения.
}

\engrus{0.5ex}{0.5ex}{
The variational formulation of this field model is similar to the appropriate formulation for Born -- Infeld nonlinear electrodynamics \cite{Chernitskii2004a,Chernitskii2012be}.
What is more, the scalar nonlinear equation under consideration has the same spherically symmetrical solution that we have in Born -- Infeld electrodynamics for zero component of the electromagnetic potential
and which is sometimes called the Born's electron.
}{
Вариационная формулировка этой полевой модели сходна с соответствующей
формулировкой электродинамики Борна -- Инфельда \cite{Chernitskii2004a,Chernitskii2012be}.
Кроме того, рассматриваемое скалярное нелинейное уравнение имеет
такое же сферически симметричное решение, которое мы имеем в электродинамике
Борна -- Инфельда для нулевой компоненты электромагнитного потенциала и которое
иногда называют борновским электроном.
}

\engrus{0.5ex}{0.5ex}{
The Lorentz transformations give the appropriate moving solitary solutions or solitons. Such solution will be called spheroidal soliton or spheron.
}{
Преобразования Лоренца дают соответствующие движущиеся уединённые решения или солитоны. Такое решение будет называться сфероидальным солитоном или сфероном.
}

\engrus{0.5ex}{0.5ex}{
As it is known the nonlinearity of the field model account for the interaction between the solitons. The appropriate methods for investigation for BI electrodynamics \cite{Chernitskii1999,Chernitskii2012be}
can be applied to the scalar field model under consideration.
}{
Как известно, нелинейность полевой модели порождает взаимодействие между солитонами. Соответствующие методы исследования для электродинамики Борна -- Инфельда \cite{Chernitskii1999,Chernitskii2012be}
могут быть применены к рассматриваемой скалярной полевой модели.
}

\engrus{0.5ex}{0.5ex}{
Here we show that the interaction between the spherons in the scalar field model under consideration looks like the electromagnetic one.
}{
Здесь мы показываем, что взаимодействие между сферонами в рассматриваемой
скалярной полевой модели выглядит как электромагнитное.
}

\engrus{3ex}{2ex}{%
\section{Extremal space-time film and energy-momentum conservation law}
\label{fmostf}
}{%
\mifeng{\addtocounter{section}{-1}}{}
\section[Экстремальная пространственно-временная плёнка]{Экстремальная пространственно-временная плёнка и закон сохранения энергии-импульса}
\mifeng{}{\label{fmostf}}
}

\engrus{0.5ex}{0.5ex}{
 The equation of the model under consideration has the following form in orthogonal coordinates
}{
Уравнение рассматриваемой модели имеет следующий вид в прямоугольных декартовых координатах
}
\begin{equation}
\label{407970191}
 \frac{\p \ffind^{\mu}}{\p x^{\mu}} = 0
\;,
\qquad
\ffind^{\mu}  \eqdef \frac{\dffun^{\mu}}{\LF}
\;,
\quad
\dffun_{\nu}  \eqdef \frac{\p\ffun}{\p x^{\nu}}
\;,
\quad
\LF = \sqrt{\left|1 + \xxx^{2}\,\metrp^{\mu\nu}\,\dffun_{\mu}\,\dffun_{\nu}\right| }
\;,
\end{equation}
\engrus{0.5ex}{0.5ex}{
\noindent
where
$\ffun$ is the scalar real field function,
$\xxx$ is the dimensional constant,
$\metrp_{\mu\nu}$ are the components of the Minkowski metric.
The Greek indices take values $\{0,1,2,3\}$.
}{
\noindent
где
$\ffun$ -- скалярная действительная полевая функция,
$\xxx$ -- размерная константа,
$\metrp_{\mu\nu}$ -- компоненты метрики Минковского.
Греческие индексы принимают значения $\{0,1,2,3\}$.
}

\engrus{0.5ex}{0.5ex}{
Customary method gives the following differential conservation law for energy-momentum tensor in the area outside of singularities:
}{
Обычный метод даёт следующий дифференциальный закон сохранения тензора энергии-импульса в области вне сингулярностей:
}
\begin{equation}
\label{583592841}
\frac{\p \EMT^{\mu\nu}}{\p x^{\mu}} = 0
\;,\qquad
\EMT^{\mu\nu}   = \frac{1}{4\pi}\left(\frac{\dffun^{\mu}\,\dffun^{\nu}}{\LF}-
\frac{\metr^{\mu\nu}}{\xxx^2}\left(\LF - 1\right)\right)
\;.
\end{equation}

\engrus{0.5ex}{0.5ex}{
Let us consider the integral from the energy-momentum differential conservation law (\ref{583592841}) over
the four-volume $\Vol$  including the three-volume $\Vols$ and the time interval
$\left[\xc^0-{\Delta x^0}/{2}\,,\;\xc^0+{\Delta x^0}/{2}\right]$.
Integration by parts in this integral gives the appropriate integral conservation law
}{
Рассмотрим интеграл от дифференциального закона сохранения (\ref{583592841}) по
четырёхмерному объёму $\Vol$, включающему трёхмерный объём $\Vols$ и временной интервал
$\left[\xc^0-{\Delta x^0}/{2}\,,\;\xc^0+{\Delta x^0}/{2}\right]$.
Интегрирование по частям в этом интеграле даёт соответствующий интегральный закон сохранения
}
 \begin{equation}
 \label{478204531}
\int\limits_{\Vol}
\frac{\p \EMT^{\mu\nu}}{\p x^{\nu}}\;\df x^{0}\df x^{1}\df x^{2}\df x^{3} = 0
\qquad\Longrightarrow\qquad\int\limits_{\Sur}\EMT^{\mu\nu}\,\dSur_\nu = 0\;,
 \end{equation}
\engrus{0.5ex}{0.5ex}{
\noindent
where
$\Sur$ is the three-dimensional boundary hypersurface for the four-volume $\Vol$,
$\dSur_\nu$ are components of the outer hypersurface element four-vector such that\mifengrus{\\}{}
$\dSur_0 = \pm \df x^1 \df x^2 \df x^3$, $\dSur_1 = \pm \df x^2 \df x^3 \df x^0$,
$\dSur_2 = \pm \df x^3 \df x^1 \df x^0$, $\dSur_3 = \pm \df x^1 \df x^2 \df x^0$.
}{
\noindent
где $\Sur$ -- трёхмерная ограничивающая гиперповерхность для четырёхмерного объёма $\Vol$,
$\dSur_\nu$ -- компоненты четырёхвектора внешнего элемента гиперповерхности такие,
что\mifengrus{}{\mifrus{\break}{}}
$\dSur_0 = \pm \df x^1 \df x^2 \df x^3$,\mifengrus{\\}{} $\dSur_1 = \pm \df x^2 \df x^3 \df x^0$,
$\dSur_2 = \pm \df x^3 \df x^1 \df x^0$, $\dSur_3 = \pm \df x^1 \df x^2 \df x^0$.
}

\engrus{0.5ex}{0.5ex}{
Let us divide the three-dimensional closed hypersurface $\Sur$ into two unclosed hypersurfaces in some coordinate system such that
$\Sur=\st{P}{\Sur}\cup\st{F}{\Sur}$:
}{
Разобъём трёхмерную замкнутую гиперповерхность $\Sur$ на две незамкнутые гиперповерхности в некоторой системе координат так, что $\Sur=\st{P}{\Sur}\cup\st{F}{\Sur}$:
}
\begin{equation}
\label{566844581}
\st{P}{\Sur} = \left.\vphantom{\Bigl|\Bigr|}\Vols\right|_{x^{0}=\xc^0-{\Delta x^0}/{2}} {}\cup{} \left.\vphantom{\Bigl|\Bigr|}\Vols\right|_{x^{0}=\xc^0+{\Delta x^0}/{2}}
\;,\quad
\st{F}{\Sur} = \Surc {}\cup{} \left[\xc^0-{\Delta x^0}/{2}\,,\;\xc^0+{\Delta x^0}/{2} \vphantom{\Bigl|\Bigr|}\right]
\;.
\end{equation}
\engrus{0.5ex}{0.5ex}{
\noindent
where $\Surc$ is the closed two-surface bounding the three-volume $\Vols$, $\st{P}{\Sur}$ will be called the momentum hypersurface and $\st{F}{\Sur}$ will be called the force hypersurface.
The three-dimensional sections for these hypersurfaces are shown on Fig.  \ref{62463995}.
}{
\noindent
где $\Surc$ -- замкнутая двумерная поверхность, ограничивающая трёхмерный объём $\Vols$, $\st{P}{\Sur}$ будет называться поверхностью импульса, а $\st{F}{\Sur}$ -- силовой поверхностью.
Трёхмерные сечения этих гиперповерхностей показаны на Рис.  \ref{62463995r}.
}
\begin{figure}[h]
\hspace{0.4pc}%
\begin{minipage}{37pc}
  {\unitlength 1mm
   \begin{picture}(120,45)
     \put(0,0){\includegraphics[width=36pc]{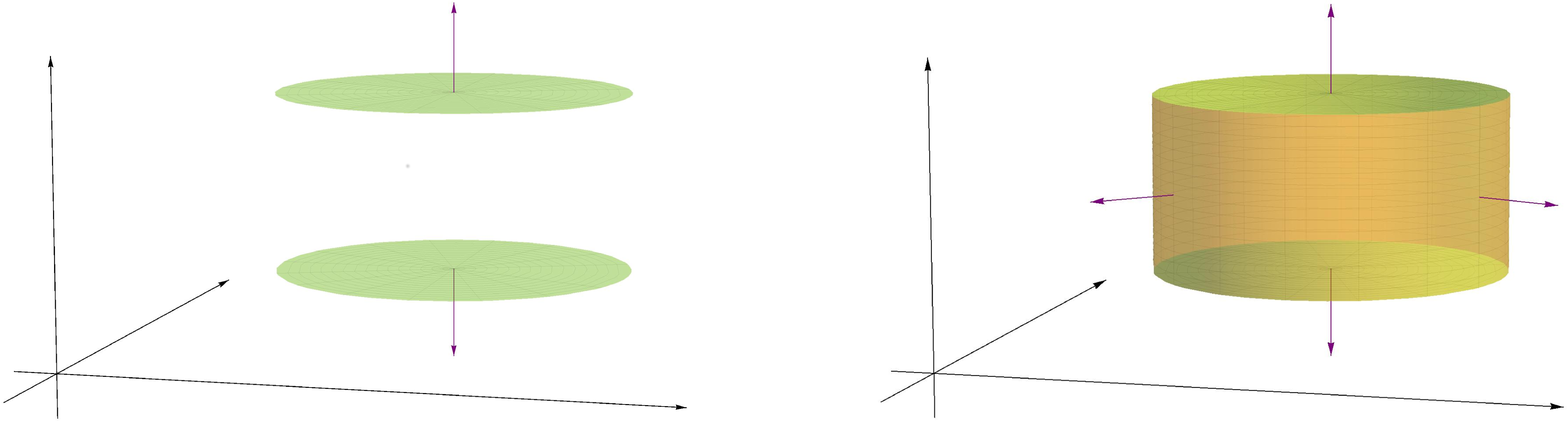}}
     \put(0,0){\small
     \put(2,33){$x^0$}
     \put(17,15){$x^2$}
     \put(62.5,3){$x^1$}
     \put(42.5,36){$\df\st{P}{\Sur}_{0}=\dVols$}
     \put(14,35){$x^{0} = \xc^0 {}+{} \frac{\Delta x^0}{2}$}
     \put(44,20.5){$x^{0} = \xc^0 {}-{} \frac{\Delta x^0}{2}$}
     \put(42.5,8){$\df\st{P}{\Sur}_{0}=-\dVols$}
     \put(30,0){(a)}
    }
    \put(76,0){\small
     \put(2,33){$x^3$}
     \put(17,15){$x^2$}
     \put(62.5,3){$x^1$}
     \put(42.5,36){$\df\st{F}{\Sur}_{3}=\df x^{1}\df x^{2}\df x^{0}$}
     \put(42.5,8){$\df\st{F}{\Sur}_{3}=-\df x^{1}\df x^{2}\df x^{0}$}
     \put(30,0){(b)}
    }
   \end{picture}
  }
\engrus{0.5ex}{0.5ex}{%
 \caption{\label{62463995} Section for $\st{P}{\Sur}$ in three-dimensional space $\{x^0,x^1,x^2\}$ (a) and
  section for  $\st{F}{\Sur}$ in three-dimensional space $\{x^1,x^2,x^3\}$ (b).}
}{%
\mifeng{\addtocounter{figure}{-1}}{}
 \caption{\label{62463995r} Сечение для $\st{P}{\Sur}$ в трёхмерном пространстве $\{x^0,x^1,x^2\}$ (a) и
  сечение для  $\st{F}{\Sur}$ в трёхмерном пространстве $\{x^1,x^2,x^3\}$ (b).}
}
\end{minipage}
\end{figure}

\engrus{0.5ex}{0.5ex}{
Thus we have the obtained integral conservation law in the following integral force dynamic law form:
}{
Таким образом имеем полученный интегральный закон сохранения в следующем виде
интегрального силового закона сохранения:
}
\begin{equation}
\label{343897031}
\Delta\EMV^\mu_{\Vols} = \bar{\Delta}\Fo^\mu_{\Surc}
\;,\qquad
\Delta\EMV^\mu_{\Vols}
\eqdef \EMV^\mu_{\Vols}
\biggr|_{x^0 = \xc^0 {}-{} \frac{\Delta x^0}{2}}^{x^0 = \xc^0 {}+{} \frac{\Delta x^0}{2}}
\;,\quad
\bar{\Delta}\Fo^\mu_{\Surc}
\eqdef \int_{\xc^0 {}-{} \frac{\Delta x^0}{2}}^{\xc^0 {}+{} \frac{\Delta x^0}{2}} \Fo^{\mu}_{\Surc}\,
\df x^0
\;,
\end{equation}
\engrus{0.5ex}{0.5ex}{
\noindent
where $\EMV^\mu_{\Vols}$ is the momentum of field into the three-dimensional volume $\Vols$, $\bar{\Delta}\Fo^\mu_{\Surc}$ is the integral force on the two-dimensional surface $\Surc$.
The momentum and the force are defined as follows:
}{
\noindent
где $\EMV^\mu_{\Vols}$ -- импульс поля в трёхмерном объёме $\Vols$, $\bar{\Delta}\Fo^\mu_{\Surc}$ -- интегральная сила на двумерной поверхности $\Surc$.
Импульс и сила определяются как:
}
 \begin{equation}
\label{34447402}
\EMV^{\mu}_{\Vols}\doteqdot \int\limits_{\Vols}\EMT^{\mu 0}\,\dVols
\;,
\qquad
\Fo^\mu_{\Surc} \eqdef -\int\limits_{\Surc} \EMT^{\mu i}\,\dSurc_i
\;.
 \end{equation}
 \engrus{0.5ex}{0.5ex}{
 \noindent
where \mbox{$\dVols = \pm\df x^1 \df x^2 \df x^3$},\mifengrus{\\}{}
\mbox{$\{\dSurc_i\}=\{\pm\df x^2\df x^3,\pm\df x^1\df x^3,\pm\df x^1\df x^2\}$} are the outer normal components of the  two-dimensional surface $\Surc$ bounded the three-dimensional volume $\Vols$.
 }{
 \noindent
где \mbox{$\dVols = \pm\df x^1 \df x^2 \df x^3$},\mifengrus{\\}{}
\mbox{$\{\dSurc_i\}=\{\pm\df x^2\df x^3,\pm\df x^1\df x^3,\pm\df x^1\df x^2\}$}
--\mifengrus{\\}{} компоненты внешней нормали двумерной поверхности $\Surc$, ограничивающей трёхмерный объём $\Vols$.
 }

\engrus{3ex}{1ex}{%
\section{Spheroidal soliton}
\label{sphsol}
}{%
\mifeng{\addtocounter{section}{-1}}{}
\section{Сфероидальный солитон}
\mifeng{}{\label{sphsol}}
}

\engrus{0.5ex}{0.5ex}{
In the case when the field function depends on the radial coordinate $r$ only we have the following simplest solution
}{
В случае, когда полевая функция зависит только от радиальной координаты $r$, имеем следующее простейшее решение
}
\begin{equation}
\label{464604551}
\frac{\p \ffun}{\p r}  = - \frac{\q}{\sqrt{r^{4} \pm \br^{4}}}
\;,
\quad
\br  \eqdef \sqrt{|\q\,\xxx|}
\quad
\Longrightarrow
\quad
\ffun  = \frac{\q}{r}\,\HypGeomF{\frac{1}{4}}{\frac{1}{2}}{\frac{5}{4}}
\!\left
(\mp\,\frac{\br^{4}}{r^{4}}
\right
)
\;,
\end{equation}
\mifengrus{}{%
\begin{figure}[h]
\begin{minipage}{11.5pc}
\vspace{-1pc}
}
\engrus{0.5ex}{0.5ex}{
\noindent
where $\HypGeomF{\alpha}{\beta}{\gamma}(z)$ is hypergeometric function,
top and bottom sings are
correspond to different signatures of metric in the model action:
$\{+,-,-,-\}$ (top) and $\{-,+,+,+\}$ (bottom).
}{
\noindent
где $\HypGeomF{\alpha}{\beta}{\gamma}(z)$ -- гипергеометрическая функция,
верхний и нижний знаки соответствуют различным сигнатурам метрики в модельном действии:
$\{+,-,-,-\}$ (верхний) and $\{-,+,+,+\}$ (нижний).
}

\mifengrus{%
\begin{figure}[h]
\hspace{0.4pc}
\begin{minipage}{37pc}}{%
\vspace{2pc}
\engrus{0.5ex}{0.5ex}{
The obtained solution (\ref{464604551}) give birth to the class of the moving soliton solutions
with the Lorentz transformations:
}{
Полученное решение (\ref{464604551}) порождает класс солитонных решений при помощи преобразований Лоренца:
}
\end{minipage}
\hspace{2pc}%
\begin{minipage}{24pc}
}
     \mifengrus{%
      {\unitlength 1.5mm
       \begin{picture}(120,32)
       \put(4,0){\includegraphics[width=34pc]{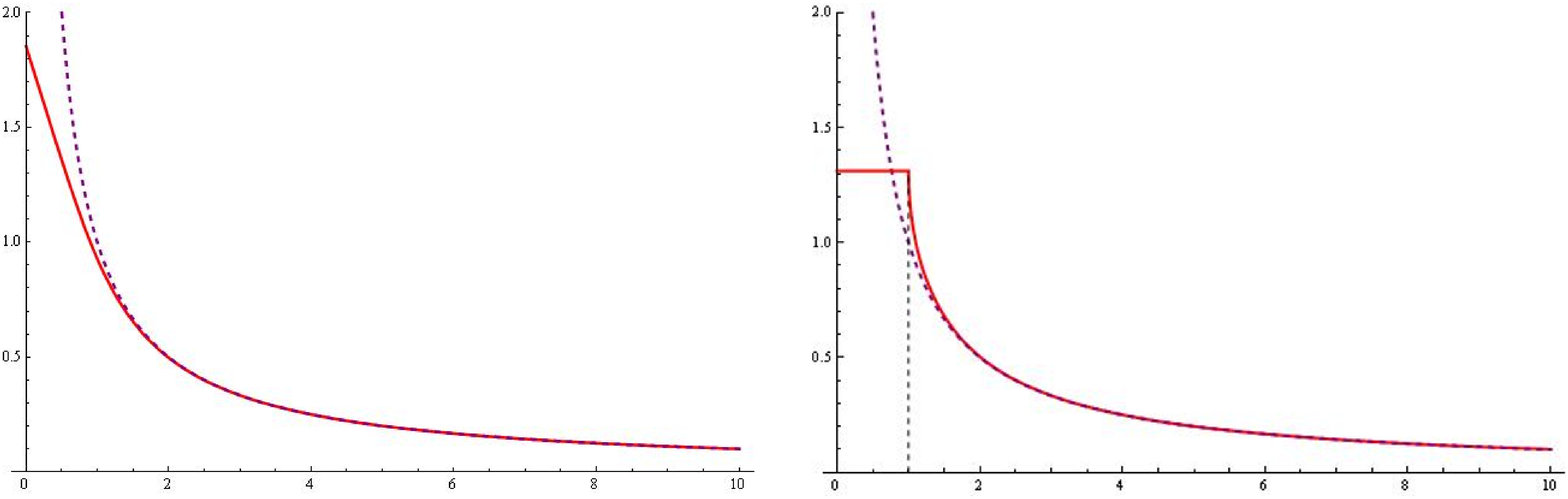}}
        \put(5,-1){\small
        \unitlength 1.2mm
        \put(-5,38){$\ffun$}
        \put(56,0){$r$}
        \put(25,-1){$(a)$}
        \put(30,25){$\q = 1$}
        \put(30,20){$\br =1$}
       }
       \put(55,-1){\small
        \unitlength 1.2mm
        \put(-5,38){$\ffun$}
        \put(56,0){$r$}
        \put(25,-1){$(b)$}
        \put(30,25){$\q = 1$}
        \put(30,20){$\br =1$}
       }
       \end{picture}
       }
     }{%
      {\unitlength 1mm
       \begin{picture}(120,30)
        \put(4,-1){\small
        \unitlength 0.9mm
        \put(-5,38){$\ffun$}
        \put(48,-1){$r$}
        \put(20,-3){$(a)$}
        \put(30,25){$\q = 1$}
        \put(30,20){$\br =1$}
       }
       \put(56,-1){\small
        \unitlength 0.9mm
        \put(-5,38){$\ffun$}
        \put(48,-1){$r$}
        \put(20,-3){$(b)$}
        \put(30,25){$\q = 1$}
        \put(30,20){$\br =1$}
       }
        \put(0,0){\includegraphics[width=24pc]{SPIN2017Proc1.eps}}
       \end{picture}
       }
       }
\engrus{0.5ex}{0.5ex}{%
 \caption{\label{35575884}The field function $\ffun (r)$ for the solution with two metric signatures: $\{+,-,-,-\}$ (a) and $\{-,+,+,+\}$ (b). The dashed lines show the function $1/r$.}
}{%
\mifeng{\addtocounter{figure}{-1}}{}
 \caption{\label{35575884r}Полевая функция $\ffun (r)$ решения при двух
 сигнатурах метрики: $\{+,-,-,-\}$ (a) и $\{-,+,+,+\}$ (b). Пунктирные
 линии показывают функцию $1/r$.}
}
\end{minipage}
\end{figure}
\mifengrus{%
\engrus{0.5ex}{0.5ex}{
The obtained solution (\ref{464604551}) give birth to the class of the moving soliton solutions
with the Lorentz transformations:
}{
Полученное решение (\ref{464604551}) порождает класс солитонных решений при помощи преобразований Лоренца:
}}{}
\begin{equation}
\label{428965661}
 \xp{x^{\mu}} = \Lor^{\mu}_{.\nu}\,x^{\nu}
 \;,\quad
 x^{\mu} = \Lor^{.\mu}_{\nu}\,\xp{x^{\nu}}
 \;,\quad
 \Lor^{\mu}_{.\nu}\,\Lor^{.\nu}_{\rho} = \Cron_{\rho}^{\mu}
\;.
\end{equation}
\engrus{0.5ex}{0.5ex}{
\noindent
Here $\{\xp{x^{\mu}}\}$ is the intrinsic coordinate system of soliton, thus we make the substitution
$r\to \xp{r} = \sqrt{\xp{x^i}\,\xp{x^i}}$ in (\ref{464604551}). The components $\Lor^{\mu}_{.\nu}$ (the point designates that the index is second) are
}{
\noindent
Здесь $\{\xp{x^{\mu}}\}$ -- собственная координатная система солитона, таким образом делаем подстановку
\mbox{$r\to \xp{r} = \sqrt{\xp{x^i}\,\xp{x^i}}$} в (\ref{464604551}). Компоненты $\Lor^{\mu}_{.\nu}$
(точка означает, что индекс -- второй) суть
}
\begin{align}
\nonumber
&\Lor^{0}_{.0} = \pu^{0} = \frac{1}{\sqrt{1 - \bxv^2}}
\;,\quad
\Lor^{0}_{.i} = \Lor^{i}_{.0} = -\pu^{i} = -\frac{\xv^{i}}{\sqrt{1 - \bxv^2}}
\;,\quad
\Lor^{i}_{.j} = \delf^{i}_{j} + \frac{\pu^{i}\,\pu^{j}}{1 + \pu^{0}}\;,
\\
&\bxv^2 \eqdef \left(\xv^{1}\right)^2 + \left(\xv^{2}\right)^2 + \left(\xv^{3}\right)^2
\;,\quad
\left|\metrp_{\mu\nu}\,\pu^{\mu}\,\pu^{\nu}\right|  = 1
\;,
\label{426156912}
\end{align}
\engrus{0.5ex}{0.5ex}{
\noindent
where $\xv^{i}$ are the components of three-dimensional velocity of the soliton, $\pu^{\mu}$ are the components of its four-dimensional velocity.
}{
\noindent
где $\xv^{i}$ -- компоненты трёхмерной скорости солитона, $\pu^{\mu}$ -- компоненты четырёхмерной скорости солитона.
}

\engrus{0.5ex}{0.5ex}{
Energy-momentum four-vector components for spheroidal soliton are
}{
Компоненты четырёхвектора энергии-импульса сфероидального солитона суть
}
\begin{equation}
\label{375089531}
\EMV^{\mu}  = \Lor^{.\mu}_{0}\,\xp{\Energy} = \pu^{\mu}\,
\mass
\;,
\end{equation}
\engrus{0ex}{0.5ex}{
\noindent
where $\mass$ \textadda{is} the energy of the soliton in its intrinsic coordinate system.
Here it should be noted that the transformation (\ref{375089531}) is true if only  we consider the moving tree-volume $\Vols$ in definition (\ref{34447402}) and transformation
of its element as the zero component of four-vector $\df\st{P}{\Sur}_{0}$.
}{
\noindent
где $\mass$ -- энергия солитона в его собственной системе координат.
Здесь надо отметить, что преобразование (\ref{375089531}) верно только, если мы рассматриваем движущийся трёхмерный объём $\Vols$ в определении (\ref{34447402}), а преобразование его элемента как нулевой компоненты четырёхвектора $\df\st{P}{\Sur}_{0}$.
}

\engrus{0.5ex}{0.5ex}{
The spheroidal soliton under consideration will be called spheron for short.
}{
Для краткости будем называть сфероидальный солитон сфероном.
}

\engrus{3ex}{2ex}{%
\section{Interacting spherons and electromagnetism}
\label{intsphaem}
}{%
\mifeng{\addtocounter{section}{-1}}{}
\section{Взаимодействующие сфероны и электромагнетизм}
\mifeng{}{\label{intsphaem}}
}

\engrus{0.5ex}{0.5ex}{
Let us consider the sum of the moving spherons which are distant from each other as the first approximation to an unknown multisoliton solution.
But we assume that the velocities of these solitons can depend on time.
To investigate the trajectories of the spherons in this first approximation we will use the integral conservation law for momentum (\ref{343897031}).
This method was used for the similar problem in nonlinear electrodynamics \cite{Chernitskii1999,Chernitskii2012be}.
}{
Рассмотрим сумму удалённых друг от друга движущихся сферонов в качестве первого приближения к неизвестному
мультисолитонному решению.
При этом предполагаем, что скорости этих солитонов могут зависеть от времени.
Для исследования траекторий сферонов в этом первом приближении будем использовать интегральный закон сохранения
импульса (\ref{343897031}).
Этот метод использовался для сходной задачи в нелинейной электродинамике \cite{Chernitskii1999,Chernitskii2012be}.
}

\engrus{0.5ex}{0.5ex}{
The calculation gives that the interaction of two spheron\textadda{s} at rest fully conforms with the electrostatic interaction of two charged point particles.
Then we can transform this static configuration to a moving reference frame with the Lorentz transformations (\ref{428965661}).
The appropriate transformation for acceleration or force is known \cite{Einstein1905aE,Pauli1958ae}. Thus for this case we obtain also the magnetic component of the force in the moving
reference frame.
}{
Вычисление даёт, что взаимодействие двух покоящихся сферонов полностью соответствует электростатическому
взаимодействию двух заряженных точечных частиц.
Затем мы можем преобразовать эту статическую конфигурацию к движущейся системе отсчёта при помощи
преобразований Лоренца (\ref{428965661}).
Соответствующее преобразование для ускорения или силы известно \cite{Einstein1905aE,Pauli1958ae}.
Таким образом в этом случае получаем также магнитную компоненту силы в движущейся системе отсчёта.
}

\engrus{0.5ex}{0.5ex}{
Let us consider the internal tensor structure of the four-vector of integral force $\bar{\Delta}\Fo^\mu_{\Surc}$ (\ref{343897031}).
}{
Рассмотрим внутреннюю тензорную структуру четырёхвектора интегральной силы $\bar{\Delta}\Fo^\mu_{\Surc}$ (\ref{343897031}).
}

\engrus{0.5ex}{0.5ex}{
Assuming that the time interval is vanishingly small ($\Delta x^{0}\to 0$)
and designating the symbols related to intrinsic coordinate system of the two spherons with point under letter we can write
}{
Предполагая, что временной интервал пренебрежимо мал ($\Delta x^{0}\to 0$)
и обозначая символы, относящиеся к собственной системе отсчёта двух сферонов, точкой под буквой, можем написать
}
\begin{equation}
\label{632072311}
\df\st{n}{\xp{\EMV}}^i   = \st{n}{\xp{\Fo}}^i\,\df \xp{x}^{0}
\;,\quad
\df\st{n}{\xp{\EMV}}^0   = 0
\;.
\end{equation}

\engrus{0.5ex}{0.5ex}{
Here we must consider $\df \xp{x}^{0}$ as zero component for four-vector with component\textadda{s} of trajectory differentials:
}{
Здесь мы должны рассматривать $\df \xp{x}^{0}$ как нулевую компоненту четырёхвектора с
дифференциалами траектории в качестве компонент:
}
\begin{equation}
\label{640498141}
 \df\xp{\xc}^{\mu} = \{\df \xp{x}^{0},0,0,0\}\quad\Longrightarrow\quad
 \df\xc^{\mu} = \{\df x^{0},\df \xc^{1},\df \xc^{2},\df \xc^{3}\}
\;.
\end{equation}

\engrus{0.5ex}{0.5ex}{
Thus to obtain the right tensor law of motion we must consider the electrostatic force components $\st{n}{\xp{\Fo}}^i$ as components of a second-rank tensor such that
}{
Таким образом, чтобы получить правильный тензорный закон движения, мы должны считать компоненты
электростатической силы $\st{n}{\xp{\Fo}}^i$ компонентами тензора второго ранга так, что
}
\begin{equation}
\df\st{n}{\EMV}^\mu   = \st{n}{\q}\,\st{n}{F}^{\mu}_{.\nu}\,\df \xc^{\nu}
\;,\quad
\st{n}{\xp{F}}^{0}_{.0} = 0
\;,\quad
\st{n}{\q}\,\st{n}{\xp{F}}^{i}_{.0} = \st{n}{\xp{\Fo}}^i
\quad\Longrightarrow\quad
\label{282090811}
\mass\, \df\st{n}{\pu}^{\mu} =  \st{n}{\q}\,\st{n}{F}^{\mu}_{.\nu}\,\st{n}{\pu}^{\nu}\,\df \st{n}{\ps}
\;,
\end{equation}
\engrus{0.5ex}{0.5ex}{
\noindent
where $\df \st{n}{\ps}^{2} {}\eqdef{} \df x^{0}\df x^{0} - \df\st{n}{\xc}^{i}\df\st{n}{\xc}^{i}$ is invariant parameter of trajectory for the $n$-th spheron.
}{
\noindent
где $\df \st{n}{\ps}^{2} {}\eqdef{} \df x^{0}\df x^{0} - \df\st{n}{\xc}^{i}\df\st{n}{\xc}^{i}$ --
инвариантный параметр траектории $n$-ого сферона.
}

\engrus{0.5ex}{0.5ex}{
Thus we have obtained that the dynamical equation for interacting spheron looks like the equation of motion for point charged particle in electromagnetic field.
But here we based on the static configuration of spherons where the electrostatic interaction was obtained.
In a conventional manner for electrostatic interaction we can introduce the appropriate electrostatic potential satisfying the Laplace's equation with a charge density as the source.
The appropriate relativistic generalization gives the four-vector electromagnetic potential components satisfying the wave equations with the appropriate four-vector current density as the source.
It is obvious that the rotating configuration with the spherons has the angular momentum or spin the dynamics of which is investigated with the help of the appropriate integral conservation law.
}{
Таким образом мы получили, что динамическое уравнение взаимодействующего сферона выглядит как уравнение движения точечной заряженной частицы в электромагнитном поле.
При этом здесь мы исходим из статической конфигурации сферонов, для которой было получено электростатическое взаимодействие.
Обычным для электростатического взаимодействия методом мы можем ввести соответствующий электростатический потенциал,
удовлетворяющий уравнению Лапласа с плотностью заряда в качестве источника.
Соответствующее релятивистское обобщение даёт компоненты четырёхвектора электромагнитного потенциала, удовлетворяющие
волновым уравнениям с соответствующей четырёхвекторной плотностью тока в качестве источника.
Очевидно, что вращающаяся конфигурация сферонов имеет момент импульса или спин, динамика которого исследуется при помощи соответствующего интегрального закона сохранения.
}

\mifengrus{\newpage}{}
\engrus{3ex}{1ex}{%
\section{Conclusions}
\label{concl}
}{%
\mifeng{\addtocounter{section}{-1}}{}
\section{Выводы}
\mifeng{}{\label{concl}}
}

\engrus{0.5ex}{0.5ex}{
Thus we have the similarity between the interacting spherons of the scalar field and the classical electrodynamics of the point charged particles.
But because the starting configuration of spherons is statical we must keep in reserve the investigation of additional effects for the interacting spherons with large relative velocities.
}{
Таким образом имеем подобие между взаимодействующими сферонами скалярного поля и классической электродинамикой
точечных заряженных частиц.
Однако, поскольку исходная конфигурация сферонов статична, мы должны оставить возможность исследования
дополнительных эффектов для взаимодействующих сферонов с большими относительными скоростями.
}

\mifengrus{\vspace{3ex}}{}
\renewcommand{\refname}{\section*{\mifengrus{References\hspace{14.6pc} Список литературы}{\mifeng{References}{Список литературы}}}}\refname
\providecommand{\newblock}{}
\mifengrus{\vspace{1ex}}{}

\end{document}